\documentclass[a4paper, twocolumn,superscriptaddress,prd,showpacs,preprintnumbers]{revtex4}
\usepackage{epsf} 
\usepackage{amsmath}
\usepackage{amssymb}
\usepackage{epsfig}
\usepackage{latexsym}
\usepackage{amsfonts}
\usepackage{dcolumn}
\usepackage{varioref}
\usepackage{ifthen}
\usepackage{graphicx}
\usepackage{amssymb,amsmath,amsthm}
\begin{document}
\parindent 0mm 
\setlength{\parskip}{\baselineskip} 
\thispagestyle{empty}
\pagenumbering{arabic} 
\setcounter{page}{1}
\mbox{ }
\preprint{UCT-TP-288/11}
\preprint{MZ-TH/11-37}
\title	{Bottom-quark mass from finite energy QCD sum rules}
\author{S. Bodenstein}
\affiliation{Centre for Theoretical \& Mathematical Physics, and Department of Physics, University of
Cape Town, Rondebosch 7700, South Africa}
\author{J. Bordes}
\affiliation{Departamento de F\'{\i}sica Te\'{o}rica,
Universitat de Valencia, and Instituto de F\'{\i}sica Corpuscular, Centro
Mixto Universitat de Valencia-CSIC}
\author{C. A. Dominguez}
\affiliation{Centre for Theoretical \& Mathematical Physics, and Department of Physics, University of
Cape Town, Rondebosch 7700, South Africa}%
\author{J. Pe\~{n}arrocha}
\affiliation{Departamento de F\'{\i}sica Te\'{o}rica,
Universitat de Valencia, and Instituto de F\'{\i}sica Corpuscular, Centro
Mixto Universitat de Valencia-CSIC}
\author{K. Schilcher}
\affiliation{Centre for Theoretical \& Mathematical Physics, and Department of Physics, University of
Cape Town, Rondebosch 7700, South Africa}
\affiliation{Institut f\"{u}r Physik, Johannes Gutenberg-Universit\"{a}t
Staudingerweg 7, D-55099 Mainz, Germany}
\date{\today}
\begin{abstract}
Finite energy QCD sum rules involving both inverse and positive moment integration kernels are employed to determine the bottom quark mass. The result obtained in the $\overline{\text {MS}}$ scheme at a reference scale of $10\, \mbox{GeV}$ is $\overline{m}_b(10\,\text{GeV})= 3623(9)\,\text{MeV}$. This value translates into a scale invariant mass $\overline{m}_b(\overline{m}_b) = 4171 (9)\, \mbox{MeV}$. This result has the lowest total uncertainty of any method, and is less sensitive to a number of systematic uncertainties that affect other QCD sum rule determinations. \\
\end{abstract}
\maketitle
%
\section{Introduction}
With the availability of new cross section data on $e^+ e^-$ annihilation into hadrons from the \emph{BABAR} collaboration \cite{babar}, the bottom quark mass was determined recently with unprecedented precision   using inverse moment QCD sum rules \cite{kuhn2009}. The result in the $\overline{\text{MS}}$ scheme at a reference scale of $10\, \mbox{GeV}$ is
\begin{equation}
\overline{m}_b(10\,\text{GeV})= 3610(16)\,\text{MeV}\,.
\end{equation}
However, as was subsequently pointed out  \cite{kuhn2010}, this result relies on the  assumption that PQCD is already valid at the end point of the  \emph{BABAR} data, i.e. $\sqrt{s} \,=\, 11.21 \, \mbox{GeV}$, where $s$ is the squared energy. This assumption might be questionable, as the prediction of PQCD for the $R$-ratio does not agree with the experimentally measured value at this point. This QCD sum rule result was also shown to depend significantly on this assumption. Hence, further reductions in the error of the bottom-quark mass using QCD sum rules will  depend on the ability to control this systematic uncertainty. One way of achieving this would be for a new experiment to extend the \emph{BABAR} measurement into a region where PQCD is unquestionably valid. In this paper we follow another approach based entirely on theory. We use  a finite energy QCD sum rule with integration kernels involving both inverse and positive powers of the energy, as employed recently to determine the charm-quark mass \cite{bodenstein}. We also exploit the freedom offered  by Cauchy's theorem to reduce the dependence of the quark mass on the above systematic uncertainty. This is achieved by using integration kernels that reduce the contributions in the region $\sqrt{s} \simeq 11.21 \, \mbox{GeV}$ to  $\sqrt{s_0}$, where there is no data and the onset of PQCD at $s=s_0$ has to be assumed. As a benefit, this procedure reduces also the continuum contribution  relative to the well known $\Upsilon$ narrow resonances.

\section{Theoretical background}
We consider the vector current correlator
\begin{eqnarray}
\Pi_{\mu\nu} (q^2) &=& i \int d^4x\,  e^{iqx} \langle 0| T(V_\mu(x) V_\nu(0))|0\rangle \nonumber\\ 
&=& (q_\mu q_\nu - q^2 g_{\mu\nu}) \Pi(q^2),
\end{eqnarray}
where $V_\mu(x) = \overline{b}(x) \gamma_\mu b(x)$, and $b(x)$ is the bottom-quark field. Cauchy's residue theorem in the complex $s$-plane ($- q^2 \equiv Q^2 \equiv s$) implies that 
\begin{eqnarray}\label{sumrule}
\int_{0}^{s_0}p(s) \frac{1}{\pi}\text{Im}\Pi(s)ds&=&-\frac{1}{2\pi i}\oint_{C(|s_0|)} p(s)\Pi(s)ds \nonumber\\
&\ +&\text{Res}[\Pi(s)\,p(s), s=0],
\end{eqnarray}
where $p(s)$ is an arbitrary Laurent polynomial, and 
\begin{equation}
\text{Im}\Pi(s)=\frac{1}{12\pi}R_b(s),
\end{equation} 
with $R_b(s)$ the standard $R$-ratio for bottom production. The power series expansion of $\Pi(s)$ for large and space-like  $s$ can be calculated in PQCD, and has the form
\begin{equation}\label{HEE}
\Pi(s)\bigl|_{\text{PQCD}}=e_{b}^{2}\sum_{n=0}\Bigl(\frac{\alpha_s(\mu^2)}{\pi}\Bigr)^n \Pi^{(n)}(s),
\end{equation}
where $e_b=2/3$ is the bottom-quark electric charge, and 
\begin{equation}
\Pi^{(n)}(s)=\sum_{i=0}\Bigl(\frac{\overline{m}_{b}^{2}}{s}\Bigr)^i \;\Pi_{i}^{(n)}.
\end{equation}
Here $\overline{m}_b\equiv\overline{m}_b(\mu)$ is the quark mass in the $\overline{\text{MS}}$ scheme, at the renormalization scale $\mu$. The  order $\mathcal{O}[\alpha_{s}^{2}(\overline{m}_{b}^{2}/s)^i]$ results for $i=1, \cdots, 6$ have been calculated in \cite{QCD1}, with new results up to
$\mathcal{O}[\alpha_{s}^{2}(\overline{m}_{b}^{2}/s)^{30}]$ obtained recently \cite{QCD1b}.
At order $\mathcal{O}[\alpha^{3}_{s}]$,  $\Pi_{0}^{(3)}$ and $\Pi_{1}^{(3)}$ are known  \cite{QCD2}, and the logarithmic terms in $\Pi_{2}^{(3)}$ may be found in  \cite{QCD3}. The constant term in $\Pi_{2}^{(3)}$ is not known exactly, but has  been estimated using Pad\'{e} approximants \cite{QCD4}, and the Mellin-Barnes transform \cite{Peris}. At order $\mathcal{O}[\alpha^{4}_{s}]$  the exact logarithmic terms in $\Pi_{0}^{(4)}$ and $\Pi_{1}^{(4)}$ were determined in \cite{QCD5}-\cite{QCD6}, whilst the constant terms are not yet known. Given that these constant terms will contribute to sum rules with kernels containing powers $s^{-1}$ and $s^0$, respectively, for  consistency we shall not include any five-loop order expressions. However, we find that if all known five-loop order terms are taken into account, the mass of the bottom-quark only changes by roughly $0.03\, \%$, which is about a tenth of the accuracy of this determination.\\
The Taylor series expansion of $\Pi(s)$ about $s=0$  is usually cast in the form
\begin{equation}
\Pi(s)\bigl|_{\text{PQCD}}=\frac{3 \,e^{2}_{b}}{16 \, \pi^2}\sum_{n\geq 0}\bar{C}_n \,z^n,
\end{equation}    
where $z\equiv s/(4\bar{m}_{b}^{2})$. The coefficients $\bar{C}_n$ can be expanded in powers of $\alpha_s(\mu)$ as
\begin{align}
\bar{C}_n=&\bar{C}_{n}^{(0)}+\frac{\alpha_s(\mu)}{\pi}(\bar{C}_{n}^{(10)}+\bar{C}_{n}^{(11)}l_m)\nonumber\\
&+\Bigl(\frac{\alpha_s(\mu)}{\pi}\Bigr)^2(\bar{C}_{n}^{(20)}+\bar{C}_{n}^{(21)}l_m+\bar{C}_{n}^{(22)}l^{2}_{m})\nonumber\\
&+\Bigl(\frac{\alpha_s(\mu)}{\pi}\Bigr)^3(\bar{C}_{n}^{(30)}+\bar{C}_{n}^{(31)}l_m+\bar{C}_{n}^{(32)}l^{2}_{m}\nonumber\\
&+\bar{C}_{n}^{(33)}l^{3}_{m})+\ldots \label{LEE}
\end{align}
where $l_m\equiv \ln(\overline{m}^{2}_b/\mu^2)$. Up to $\mathcal{O}(\alpha_{s}^{2})$, the coefficients up to $n=30$ of $\bar{C}_n$ are known \cite{QCD8}-\cite{QCD9}. 
There is also a  sub-leading contribution of order ${\cal{O}}(\alpha_s^2 \; (\overline{m}_c/\overline{m}_b)^2)$ \cite{Corcella} affecting the coefficient $\overline{C}_n^{(20)}$ in Eq.\eqref{LEE}, as well as QED corrections. The former contributes around  $- 1.0\, \mbox{MeV}$, and the latter roughly $- 2.0 \,\mbox{MeV}$ to the result for $\overline{m}_b(10 \, \mbox{GeV})$. 
Finally, there is  a non-perturbative contribution to $\Pi(s)$ from the gluon condensate, but it has been  found to be completely negligible \cite{kuhn2007}. We fully agree, and thus confirm this result. 
For the strong running coupling we use the Particle Data Group \cite{PDG} value  $\alpha_s(m_{Z})=0.1184(7)$, which corresponds to $\alpha_s(10\,\text{GeV})=0.1792(16)$.
\section{Experimental Input}
\begin{figure}
[ht]
\begin{center}
\includegraphics[scale=0.67]{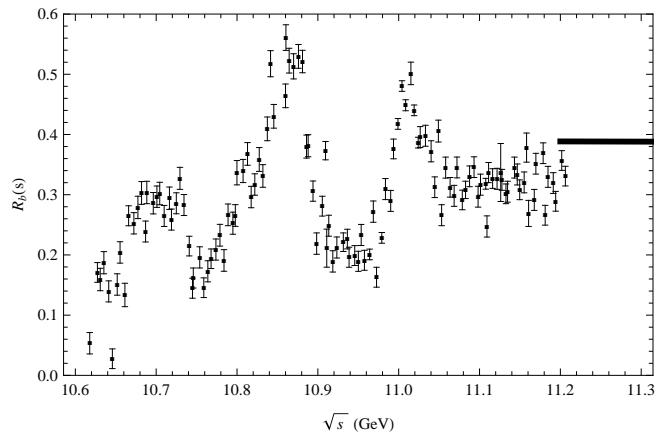}
\caption{The corrected \emph{BABAR} data \cite{babar} and the PQCD prediction (solid black line) obtained using \texttt{Rhad} \cite{rhad}.}\label{Fig:data}
\end{center}
\end{figure}

In order to  evaluate the left-hand side of Eq.\eqref{sumrule} one needs to use experimental input. First, there are the four narrow $\Upsilon$-resonances, and we calculate their contribution to Eq.\eqref{sumrule} using the zero-width approximation
\begin{equation}
R_{b}^{\text{res}}=\sum_{i}\frac{9 \,\pi \,M_i \,\Gamma_i}{\alpha_{\text{EM}}^{2}(s)}\, \delta(s-M_{i}^2)\,,
\end{equation}
where $i=1, \cdots, 4$, corresponding to $\Upsilon(1S)$, $\Upsilon(2S)$, $\Upsilon(3S)$, and $\Upsilon(4S)$. We use the masses and widths from the Particle Data Group \cite{PDG}. The widths are $\Gamma_{\Upsilon(1S)}=1.340(18)\,\text{keV}$, $\Gamma_{\Upsilon(2S)}=0.612(11)\,\text{keV}$, $\Gamma_{\Upsilon(3S)}=0.443(8)\,\text{keV}$ and $\Gamma_{\Upsilon(4S)}=0.272(29)\,\text{keV}$. Given that the widths of the $\Upsilon(1S),\Upsilon(2S)$ and $\Upsilon(3S)$ were obtained at the same experimental facility, we will assume  their uncertainties to be correlated. The masses are $M_{\Upsilon(1S)}=9.46030(26)\,\text{GeV}$, $M_{\Upsilon(2S)}=10.02326(31)\,\text{GeV}$, $M_{\Upsilon(3S)}=10.3552(5)\,\text{GeV}$, and $M_{\Upsilon(4S)}=10.5794(12)\,\text{GeV}$. Finally, we use the effective electromagnetic couplings from \cite{kuhn2007}.  The \emph{BABAR} Collaboration \cite{babar} has performed direct measurements of $R_b$ in the {\it continuum threshold} region between $10.62\,\text{GeV}$ and $11.21\,\text{GeV}$.  There is also data on the full ratio R in the bottom-quark region by the  CLEO Collaboration \cite{cleo} dating back to 1985. Subsequently, a later CLEO measurement in 1998 \cite{cleo2}, at a single energy, $s \simeq 10.53 \;\mbox{GeV}^2$, gives a total R-ratio roughly 30\% lower than the 1985 data in this region. Since this discrepancy remains unresolved we shall use here only the \emph{BABAR} data.
As was pointed out in \cite{kuhn2009}, these \emph{BABAR} data cannot be used directly in sum-rules, such as e.g. Eq.\eqref{sumrule}, for the following reasons. First, the initial-state radiation and the radiative tail of the $\Upsilon_{4S}$ resonance must be removed. Second, the vacuum polarization contribution must be taken into account. We follow this procedure, as detailed in \cite{kuhn2009}, to correct the \emph{BABAR} data with results shown in Fig. \ref{Fig:data}.\\ The high-energy expansion of $\Pi(s)$, given in Eq.\eqref{HEE}, is only formally guaranteed to converge above $\sqrt{s}= 4 \,\overline{m}_b (\mu) \approx 15\,\text{GeV}$, due to non-planar diagrams having cuts starting there. Above this value the high energy expansion is an almost perfect approximation to the full analytic PQCD result \cite{rhad}. Therefore, we shall always choose $\sqrt{s_0}> 4\, \overline{m}_b (\mu)$ in Eq.\eqref{sumrule} so that it is  safe to use the high energy expansion of $\Pi(s)$ in the contour integral.  Between the end point of the data ($\sqrt{s}=11.21\,\text{GeV}$) and $\sqrt{s_0}> 4\, \overline{m}_b (\mu)$, we will use the best available PQCD prediction of $R_b(s)$, obtained from the Fortran program \texttt{Rhad} \cite{rhad}. We  consider this as {\it data input}, even though it stems from theory. The \texttt{Rhad} \cite{rhad} prediction of $R_{b}(s)$ is shown in Fig. \ref{Fig:data}.\\
\begin{figure}
\begin{center}
\includegraphics[scale=0.67]{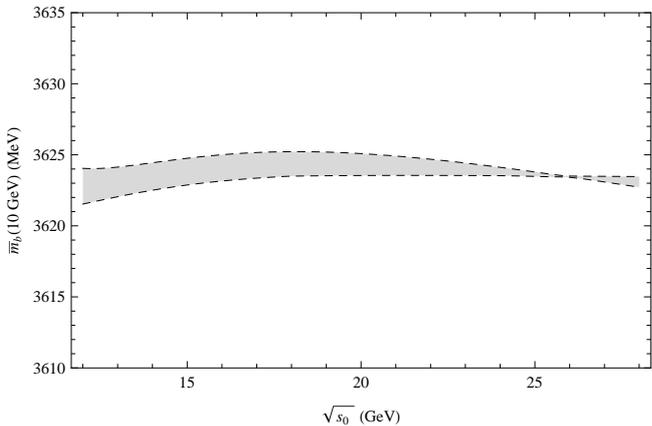}
\caption{The values of $\overline{m}_b(10\,\mbox{GeV})$, obtained for different values of $s_0$ and using the 10 different kernels in the class $\mathcal{P}_{3}^{(i,j,k)}(s_0,s)$. All results lie within the shaded region.}\label{Fig:stab1}
\end{center}
\end{figure}
\begin{table*}[ht!]
\begin{ruledtabular}
\begin{tabular}{lccccccccc}
& &  & \multicolumn{4}{c}{Uncertainties (MeV)} &  \multicolumn{3}{c}{\textbf{Options A, B, C} (MeV)}\\
\cline{4-7}
\cline{8-10}
\noalign{\smallskip}
$p(s)$ 	& $\overline{m}_b(10\,\mbox{GeV})$ & $\sqrt{s_0}\,(\text{GeV})$  & $\Delta\text{EXP.}$ & 	$\Delta \alpha_s$  &	 $\Delta \mu$			&	 $\Delta$TOTAL	  &  $\Delta \textbf{A}$  &  $\Delta \textbf{B}$   & $\Delta \textbf{C}$    \\
\hline
\noalign{\smallskip}
$s^{-3}$ 			&  	3612  						&  $\infty$		&		9	 		&		4	 	&		1	 &	10		 & 		20	 &	-17 & 16 \\
$s^{-4}$ 			&  	3622  						&  $\infty$		&		7	 		&		5	 	&		10	 &	13		 & 		12	 &	-12 & 8 \\ \\
$\mathcal{P}_{3}^{(-3,-1,0)}(s_0,s)$ 	&  	 3623 		& 16			&   6		 &		6	 	&		2	 &	9		 & 		1	 &	-6 & 0 \\
$\mathcal{P}_{3}^{(-3,-1,1)}(s_0,s)$ 	&  	 3623 		& 16			&   6		 &		6	 	&		2	 &	9		 & 		2	 &	-7 & 0 \\
$\mathcal{P}_{3}^{(-3,0,1)}(s_0,s)$ 	&  	 3624 		& 16			&   7		 &		6	 	&		2	 &	9		 & 		2	 &	-7 & 0 \\
$\mathcal{P}_{3}^{(-1,0,1)}(s_0,s)$ 	&  	 3625 		& 16			&   8		 &		5	 	&		4	 &	10		& 	4	 &	-12 & 0 \\ \\
$\mathcal{P}_{4}^{(-3,-1,0,1)}(s_0,s)$ 	&  	 3623 		& 20			&   6		 &		6	 	&		3	 &	9		& 	0	 &	-4 & 0 \\
\end{tabular}\caption{Results for $\overline{m}_b(10\,\mbox{GeV})$ using kernels $p(s)$ selected for producing the lowest uncertainty. Results from the kernels $p(s)=s^{-3}$ and $p(s)=s^{-4}$ used in \cite{kuhn2009}-\cite{kuhn2010} are given here for comparison. The errors are from experiment ($\Delta\text{EXP.}$), the strong coupling ($\Delta \alpha_s$) and variation of the renormalization scale by $\pm \, 5 \, \mbox{GeV}$ around $\mu = 10\, \mbox{GeV}$ ($\Delta \mu$). These sources were added in quadrature to give the total uncertainty ($\Delta$TOTAL). The option uncertainties $\Delta \textbf{A}$, $\Delta \textbf{B}$ and $\Delta \textbf{C}$ are the differences between $\overline{m}_b(10\,\mbox{GeV})$ obtained with and without  \textbf{Option A},  \textbf{B},  or \textbf{C}. As in  \cite{kuhn2009}-\cite{kuhn2010} these are not added to the total uncertainty, and are listed only for comparison purposes.}\label{Tab:results}
\end{ruledtabular}
\end{table*}
The first  uncertainties affecting the bottom-quark mass are due to the uncertainty in the strong coupling $\alpha_s$ $(\Delta \alpha_s)$, the uncertainty in the experimental data ($\Delta \text{EXP})$, and our limited knowledge of PQCD ($\Delta \mu)$.  The latter was estimated by varying the renormalization scale $\mu=10\,\text{GeV}$ by $\pm \,5 \,\text{GeV}$, running the mass calculated at this scale back to $\mu=10\,\text{GeV}$ and then taking the maximum difference. 
The second set are systematic uncertainties stemming from the fact that the PQCD prediction for $R_b(s)$ does not agree with the experimentally determined values at the end point of the data  ($\sqrt{s}=11.21\,\text{GeV}$), as can be seen from Fig. \ref{Fig:data}. Two possibilities for this discrepancy were considered in \cite{kuhn2010}. \textbf{Option A}: The \emph{BABAR} data are correct, but PQCD only starts at higher energies, say at $\sqrt{s}=13\,\text{GeV}$. Use then a linear interpolation between $R_{b}^{\text{EXP}}(11.21\,\text{GeV})=0.32$ and $R_{b}^{\text{PQCD}}(13\,\text{GeV})=0.377$, rather than the prediction from \texttt{Rhad}. \textbf{Option B}: The PQCD prediction from \texttt{Rhad} is correct, but the \emph{BABAR} data are incorrect, perhaps affected by an unreported systematic error. In this case multiply all the data by a factor of 1.21 to make the data consistent with PQCD.  In addition to these two options, we wish to consider a third possibility. \textbf{Option C}: The \emph{BABAR} data are correct, and PQCD starts at $\sqrt{s}=11.21\,\text{GeV}$. However, the PQCD prediction of \texttt{Rhad} is incorrect. The motivation for this option is that the exact analytical form of $R_{b}^{\text{PQCD}}$ is only known up to   one-loop level. At order $\mathcal{O}(\alpha_{s}^{2})$ already the full analytic result has to be reconstructed using Pad\`e approximants to patch together information about $\Pi(s)$ obtained at $\sqrt{s}=0, \sqrt{s}= 2 \,\overline{m}_b (\mu)$ and $\sqrt{s} \rightarrow - \infty$. Both the Pad\`e method, and the reliance on PQCD results obtained at threshold ($\sqrt{s}= 2 \,\overline{m}_b (\mu)$) could introduce unaccounted systematic errors. As a measure of the dependence of the method on the prediction of $R_{b}^{\text{PQCD}}(s)$ up to $s_0$ (chosen to be large enough so that the high energy expansion becomes a rigorous prediction), we  use $R_{b}^{\text{PQCD}}(s)$ calculated using the high energy expansion. The prediction of $R_{b}^{\text{PQCD}}$ at $\sqrt{s} = 11.21\,\text{GeV}$ using the high energy expansion is also closer to experiment than the prediction obtained using \texttt{Rhad}. 
\section{Choice of integration kernels}
To minimize the dependence of results for the bottom-quark mass on \textbf{Option A} and \textbf{Option C}, the contribution from the region $\sqrt{s} \equiv \sqrt{s^*} \equiv 11.21\,\text{GeV}$ to $\sqrt{s_0}$ should be quenched. This can be achieved by borrowing from the method of \cite{schilcher}, where a Legendre polynomial was used to minimize the contribution of the then poorly known {\it continuum threshold} region. We choose here a Legendre-type Laurent polynomial, i.e. we  consider linear combinations of powers of $s$ chosen from the set $\mathcal{S}=\{s^{-3},s^{-2},s^{-1},1,s\}$. Inverse powers higher than $s^{-3}$ lead to a deterioration of the convergence of PQCD, introducing large uncertainties from changes in the renormalization scale $\mu$ and the strong coupling $\alpha_s$ (see also \footnote{The $\overline{\text{MS}}$-bar mass becomes an inappropriate mass scheme when using such low moments, and an alternative mass scheme should be used. See for example \cite{signer}.}). We only use positive powers up to $s^1$, as higher powers emphasize unknown $\mathcal{O}(\alpha_{s}^{3})$ terms in the high energy expansion. The optimal order of the Legendre-type Laurent polynomial was found to be $3$ or $4$. First, let us consider the order $3$ case and let
\begin{equation}\label{intkernel}
 p(s) \equiv \mathcal{P}_{3}^{(i,j,k)}(s,s_0)=A(s^{i}+B s^j+C s^k)\;, 
\end{equation} 
subject to the global constraint
\begin{equation}\label{constraint}
\int_{s^*}^{s_0} \mathcal{P}_{3}^{(i,j,k)}(s,s_0)\; s^{-n}\;ds= 0,
\end{equation} 
where $n\in \{0,1\}$, $i,j,k\in \{-3,-2,-1,0,1\}$, and $i,j,k$ are all different. The above constraint determines the constants $B$ and $C$. The constant $A$ is an arbitrary overall normalization which cancels out in the sum rule Eq.\eqref{sumrule}. The reason for the presence of the integrand $s^{-n}$ above is that the behavior of $R_b(s)$ in the region to be quenched resembles a monotonically decreasing logarithmic function. Hence, an inverse power of $s$ optimizes the quenching.
As an example, taking $s_0= (16\,\text{GeV})^2$ (and $A=1$) we find
\begin{eqnarray}
\mathcal{P}_{3}^{(-3,-1,0)}(s,s_0)&=& s^{-3} - (1.02\times 10^{-4} \;\mbox{GeV}^{-4}) \;s^{-1} \nonumber\\
&+&  3.70\times 10^{-7}\; \mbox{GeV}^{-6} \;,
\end{eqnarray}
with $s$ in units of $\mbox{GeV}^2$.
There are  ten different kernels $\mathcal{P}_{3}^{(i,j,k)}$, and  the spread of values obtained for $\overline{m}_b$ using this set of different kernels will be used as a consistency check on the method. Outside the interval $s\in[s^*,s_0]$, $\mathcal{P}_{3}^{(i,j,k)}(s,s_0)$ will blow-up, which leads to a suppression of the {\it continuum threshold} region relative to the well measured $\Upsilon$-resonances. This will minimize the dependence of the results on \textbf{Option B}. Hence, this kernel  minimizes all three sources of systematic uncertainty.
The fourth-order Laurent polynomial $\mathcal{P}_{4}^{(i,j,k,r)}(s,s_0)$ is also defined by the constraint Eq.\eqref{constraint}, but with $n\in \{0,1,2\}$. There are also five different kernels $\mathcal{P}_{4}^{(i,j,k,r)}(s,s_0)$. In general, the higher the order  $n$ of $\mathcal{P}_{n}$, the better the control over the systematic errors. However, the price to pay is a reduction in the rate of convergence of PQCD, though this convergence can be improved by increasing $s_0$. In the Appendix we give explicit expressions for the various kernels used in table I.
\section{Results and Conclusions}
We considered a total of 15 different kernels $p(s)$ used in Eq.\eqref{sumrule}, 10 from the class of kernels $\mathcal{P}_{3}^{(i,j,k)}(s,s_0)$ and 5 from the class $\mathcal{P}_{4}^{(i,j,k,r)}(s,s_0)$. All these are similarly constructed (i.e they obey Eq.\eqref{constraint}), and hence have a similar ability to reduce the dependence of the bottom-quark mass on \textbf{Options A, B, C}. They do, however, place very different emphasis on theory. In particular, if say  $\mathcal{P}_{3}^{(i,j,k)}(s,s_0)$ only included inverse powers of $s$, then almost the entire right hand side of Eq.\eqref{sumrule} would emanate from the residue, and hence from the low energy expansion of PQCD. If however $\mathcal{P}_{3}^{(i,j,k)}(s,s_0)$ were composed of only positive powers of $s$, then only the high energy expansion of PQCD would enter the right hand side of Eq.\eqref{sumrule}. Different kernels can therefore lead to significantly different dependencies on the renormalization scale $\mu$. Our philosophy is to choose those kernels producing the lowest total uncertainty. The results from these are displayed in Table \ref{Tab:results}. We also plot in Fig.2 the range of values for $\overline{m}_b(10\,\text{GeV})$ obtained using all of the  10 kernels in the class $\mathcal{P}_{3}^{(i,j,k)}(s,s_0)$, as a function of  $s_0$. Remarkably, between $12\,\text{GeV}<\sqrt{s_0}<28\,\text{GeV}$, all of the masses obtained using all 10 kernels from the class $\mathcal{P}_{3}^{(-3,-1,0)}(s,s_0)$ fall in the range $3621\,\text{MeV}\leq\bar{m}_b(10\,\text{GeV})\leq 3625\,\text{MeV}$. Our method gives a consistent result even in the region $\sqrt{s_0}< 4 \overline{m}_b (\mu)\approx 15\,\text{GeV}$ where the high-energy expansion used in the contour integral in Eq.\eqref{sumrule} is not guaranteed to converge. Using, rather, the 5 kernels in the class $\mathcal{P}_{4}^{(i,j,k,r)}(s,s_0)$, and varying $s_0$ in the range $18\,\text{GeV}<\sqrt{s_0}<70\,\text{GeV}$, all of the masses thus obtained lie in the interval $3620\,\text{MeV}\leq\bar{m}_b(10\,\text{GeV})\leq 3626\,\text{MeV}$. These results show a great insensitivity of our method on the  parameter $s_0$, and also on which powers of $s$ are used to construct $\mathcal{P}_{3}^{(i,j,k)}(s,s_0)$ and $\mathcal{P}_{4}^{(i,j,k,r)}(s,s_0)$. This in turn demonstrates the consistency between the high and low energy expansions of PQCD. \\  For our final result we choose the optimal kernel $\mathcal{P}_{3}^{(-3,-1,0)}(s_0,s)$ to obtain
\begin{equation}
\overline{m}_b(10\,\text{GeV})= 3623(9)\,\text{MeV} \;,  \label{final}
\end{equation}
\begin{equation}
\overline{m}_b(\overline{m}_b)= 4171(9)\,\text{MeV} \;.  \label{final2}
\end{equation}
This result  is fully consistent with the latest lattice value $\overline{m}_b(10\,\text{GeV})= 3617(25)\,\text{MeV}$ \cite{lattice}. It is also consistent with a previous QCD sum rule precision determination \cite{kuhn2009}-\cite{kuhn2010} giving $\overline{m}_b(10\,\text{GeV})= 3610(16)\,\text{MeV}$. Apart from our novel QCD sum rule approach, the
inputs in the latter are almost identical to ours, with the exception of their use of kernels of the form $p(s)=s^{-n}$, $n\in \{2,3,4,5\}$, and  the use of a value of the strong coupling with a  larger uncertainty. Their final result was obtained using $p(s)=s^{-3}$, which can be seen from Table \ref{Tab:results} as being far more sensitive to possible systematic uncertainties arising from \textbf{Options A, B, C}. They also determined $\overline{m}_b$ using $p(s)=s^{-4}$, for which they obtained $\overline{m}_b(10\,\text{GeV})= 3619(18)\,\text{MeV}$. This value is closer to our result, which may not be surprising given that it is less sensitive to \textbf{Options A, B, C} than $p(s)=s^{-3}$, although not as insensitive as using our kernels.\\ 
In conclusion, we have discussed here a finite energy QCD sum rule method with  integration kernels involving inverse and positive powers of the squared energy. The result for the bottom-quark mass has a lower total uncertainty, and is far less sensitive than the popular inverse moment method to the three systematic uncertainties identified earlier, i.e. \textbf{Options A, B, C}. It should be appreciated from Table I that the results Eqs.\eqref{final}-\eqref{final2} are independent of the PQCD prediction from \texttt{Rhad} in the region between $\sqrt{s} \simeq 11.21 \, \mbox{GeV}$ and $\sqrt{s} = 4 \overline{m}_b(\mu)$. 
\section{Appendix}
Up to an overall constant, the integration kernels $\mathcal{P}_{n}(s,s_0)$ can be obtained from Eq.(11). For completeness we list below the explicit expressions for all the polynomials used in Table I, at the corresponding values of $s_0$. First, for $s_0=(16\,\text{GeV})^2$ 
\begin{eqnarray}
\mathcal{P}_{3}^{(-3,-1,0)}(s,s_0) &=& s^{-3}-(1.015\times 10^{-4}\,\text{GeV}^{-4})\,s^{-1}\nonumber\\
&+& 3.694\times 10^{-7}\,\text{GeV}^{-6}\;, 
\end{eqnarray}

\begin{eqnarray}
\mathcal{P}_{3}^{(-3,-1,1)}(s,s_0) &=& s^{-3}-(6.875\times 10^{-5}\,\text{GeV}^{-4})\,s^{-1}\nonumber\\
&+&(1.000\times 10^{-9}\,\text{GeV}^{-8})s \;, 
\end{eqnarray}

\begin{eqnarray}
\mathcal{P}_{3}^{(-3,0,1)}(s,s_0) &=& s^{-3}-7.767\times 10^{-7}\,\text{GeV}^{-6}\nonumber\\
&+&(3.103\times 10^{-9}\,\text{GeV}^{-8})s\;,
\end{eqnarray}

\begin{eqnarray}
\mathcal{P}_{3}^{(-1,0,1)}(s,s_0) &=& s^{-1}-0.01129\,\text{GeV}^{-2}\nonumber\\
&+&(3.059\times 10^{-5}\,\text{GeV}^{-4})s \;.
\end{eqnarray}
\vspace{.2cm}

Next, for $s_0=(20\,\text{GeV})^2$ 

\begin{eqnarray}
&&\mathcal{P}_{3}^{(-3,-1,0,1)}(s,s_0) = s^{-3}-(1.4668\times 10^{-4}\,\text{GeV}^{-4})\,s^{-1}\nonumber\\
&+& 8.781\times 10^{-7}\,\text{GeV}^{-6}-(1.381\times 10^{-9}\,\text{GeV}^{-8})\,s \;.
\end{eqnarray}

\section{Acknowledgements}
This work was supported in part by the National Research Foundation (South Africa) and by the Alexander von Humboldt Foundation (Germany). The authors thank Hubert Spiesberger for discussions on the data, and one of us (SB) wishes to thank C. Sturm for helpful correspondence.


\begin{thebibliography}{99}

\bibitem{babar} B. Aubert {\it{et al}}., Phys. Rev. Lett. {\bf 102}, 012001 (2009).

\bibitem{kuhn2009}  K. G. Chetyrkin {\it{et al.}}, Phys. Rev. D {\bf 80}, 074010 (2009).

\bibitem{kuhn2010}  K. G. Chetyrkin {\it{et al.}}, arXiv:1010.6157v2 (2010).

\bibitem{bodenstein}  S. Bodenstein, {\it{et al.}}, Phys. Rev. D {\bf 83}, 074014 (2011).

\bibitem{QCD1} K. G. Chetyrkin, R. Harlander, J. H. K\"{u}hn, and M. Steinhauser,  Nucl. Phys. B {\bf 503}, 339 (1997).

\bibitem{QCD1b} A. Maier, and P. Marquard, arXiv: 1110.558. 
 
\bibitem{QCD2}P. A. Baikov, K. G. Chetyrkin, and J. H. K\"{u}hn, Nucl. Phys. B (Proc. Suppl.) {\bf 189}, 49 (2009).

\bibitem{QCD3}K. G. Chetyrkin, R. Harlander, J. H. K\"{u}hn, Nucl. Phys. B {\bf 586}, 56 (2000).

\bibitem{QCD4}Y. Kiyo, A. Maier, P. Maierh\"{o}fer, and P. Marquard, Nucl. Phys. B {\bf 823}, 269 (2009). 

\bibitem{Peris} D. Greynat, and S. Peris, Phys. Rev. D {\bf 82} 034030 (2010). 

\bibitem{QCD5} P. A. Baikov, K. G. Chetyrkin, and J. H. K\"{u}hn, Phys. Rev. Lett. {\bf 101}, 012002 (2008).

\bibitem{QCD6} P. A. Baikov, K. G. Chetyrkin, and J. H. K\"{u}hn, Nucl. Phys. B (Proc. Suppl.) {\bf 135}, 243 (2004).

\bibitem{QCD8} R. Boughezal, M. Czakon, and T. Schutzmeier, Phys. Rev. D {\bf 74}, 074006 (2006); Nucl. Phys. B (Proc. Suppl.) {\bf 160}, 164 (2006).

\bibitem{QCD9}A. Maier, P. Maier\"{o}fer, and P. Marquard, Nucl. Phys. B {\bf 797}, 218 (2008); Phys. Lett. B {\bf 669}, 88 (2008).

\bibitem{Corcella} G. Corcella and A. H. Hoang, Phys. Lett. B {\bf 554}, 133 (2003).


\bibitem{kuhn2007} J. H. K\"{u}hn, M. Steinhauser, and C. Sturm, Nucl. Phys. B {\bf 778}, 192 (2007).

\bibitem{PDG} K. Nakamura {\it{et al.}}, Particle Data Group, J. Phys. G {\bf 37}, 075021 (2010).

\bibitem{cleo} D. Besson {\it{et al}.}, Phys. Rev. Lett. {\bf 54}, 381 (1985).

\bibitem{cleo2} R. Ammar {\it{et al.}}, Phys. Rev. D {\bf{57}}, 1350 (1998).

\bibitem{rhad} R.V. Harlander and M. Steinhauser, Comput. Phys. Commun. {\bf 153}, 244 (2003).

\bibitem{schilcher}  J. Bordes, J. Pe\~narrocha, K. Schilcher, Phys. Lett. B {\bf 562}, 81 (2003).

\bibitem{signer}  A. Pineda, A. Signer, Phys. Rev. D {\bf 73}, 111501 (2006).

\bibitem{lattice}  C. McNeile {\it{et al}.}, Phys. Rev. D {\bf 82}, 034512 (2010).

\end{thebibliography}
\end{document}